\documentclass[graybox]{svmult}

\usepackage{mathptmx}       
\usepackage{helvet}         
\usepackage{courier}        
\usepackage{type1cm}        
\usepackage{makeidx}         
\usepackage{graphicx}        
\usepackage{multicol}        
\usepackage[bottom]{footmisc} 
\usepackage{natbib}
\newcommand{\feh}{\ensuremath{[{\rm Fe}/{\rm H}]}} 

\newcommand{\rcz}{{\rm R_{CZ}}}
\newcommand{\ys}{{\rm Y_S}}
\newcommand{\dc}{\left<\delta c/c\right>}
\newcommand{\drho}{\left<\delta \rho/\rho\right>}
\newcommand{\zxo}{\rm \left(Z/X  \right)_{\odot}}

\makeindex      
\begin{document}

\title*{Solar abundance problem}
%
%
\author{Maria Bergemann and Aldo Serenelli}
%
%
\institute{M. Bergemann \at Institute of Astronomy, University of Cambridge,
Madingley Road, CB3 0HA, Cambridge, UK\\\email{mbergema@ast.cam.ac.uk} \and  A.
Serenelli \at Institut  de Ci\`encies de  l'Espai,  Campus   UAB, Fac.  de 
Ci\`encies,  Bellaterra,   08193,  Spain\\\email{aldos@ice.csic.es}} 
%
%
\maketitle

\abstract*{}
\abstract{The chemical composition of the Sun is among the most important
quantities in astrophysics. Solar abundances are needed for modelling stellar
atmospheres, stellar structure and evolution, population synthesis, and galaxies
as a whole. The solar abundance problem refers to the conflict of observed data
from helioseismology and the predictions made by stellar interior models for
the Sun, if these models use the newest solar chemical composition obtained with
3D and NLTE models of radiative transfer. Here we take a close look at the
problem from observational and theoretical perspective. We also provide a list
of possible solutions, which have yet to be tested.}

\section{Introduction}
\label{sec:1}

Until recently, we thought we understand the Sun very well: its surface
temperature, surface  pressure, age and mass,  interior physical properties,
abundances of different chemical elements.  After all, most of these parameters
can be determined  by very  precise direct  methods: solar  effective 
temperature is known to  an astonishing accuracy of  $0.01 \%$, as measured 
from the radiant bolometric flux;  neutrino fluxes provide  the temperature in
the  solar   core;  solar   age  is  obtained   from  isotopic  ratios   in 
meteorites; helioseismolgy  - the  analysis of  propagation  of acoustic  waves
in the solar interior -  gives accurately  the  depth  of  the convective  zone 
and  surface  helium abundance. It turned out, however, that what  is still not
known  exactly is the solar  chemical composition. The main reasons for this gap
in our knowledge of the Sun will be discussed in this lecture.

The first  paper giving  a reference set of  the solar  abundances for many
chemical elements - standard  solar  composition  (SSC),  appeared  about a
century  ago \citep{1929ApJ....70...11R}. However,  only  a  few  decades 
later,  when  computer  power increased enough to  run complex numerical 
algorithms, this dataset acquired  its main value. The abundances were rightly
plugged into a  variety of astrophysical models. The first major application of 
SSC was found in  the standard solar models  (SSM), which  predict evolution  of
the  Sun from  its  formation till present. SSC went  into models of stellar
evolution,  stellar populations, and galaxies, becoming a ruler for measuring 
how dissimilar from {\it the  Sun} are  other  cosmic  objects. Highly accurate 
solar  abundance distributions  are nowadays  needed   for  research  into the
physics of Galactic formation and evolution \citep[e.g.][]{2006A&A...451.1065G, 
2013NewAR..57...80F}, to search  for solar twins, i.e. stars very similar to the
 Sun and thus potentially hosting earth-like planets
\citep{2012A&A...543A..29M}. The Sun has also traditionally been used as a
laboratory for particle physics, particularly for setting constraints on the
properties of dark matter candidates such as axions, using the sensitivity of
helio-seismic probes of the solar structure \citep{1999APh....10..353S}. Also in
this case, highly accurate chemical composition of  the Sun is needed:
abundances in the solar interior determine the radiative opacities and affect
the interaction between dark matter and baryons. For example, for certain dark
matter candidate particles, interaction with baryons depends strongly on the
properties of nuclei -charge, spin- and a detailed knowledge of the chemical
composition and profiles in the solar interior is necessary. Another example is
that of non-annihilating  dark matter particles, which can  strongly modify the
energy transport in the solar interior.
\begin{figure}[b]
\begin{center}
\includegraphics[scale=.35]{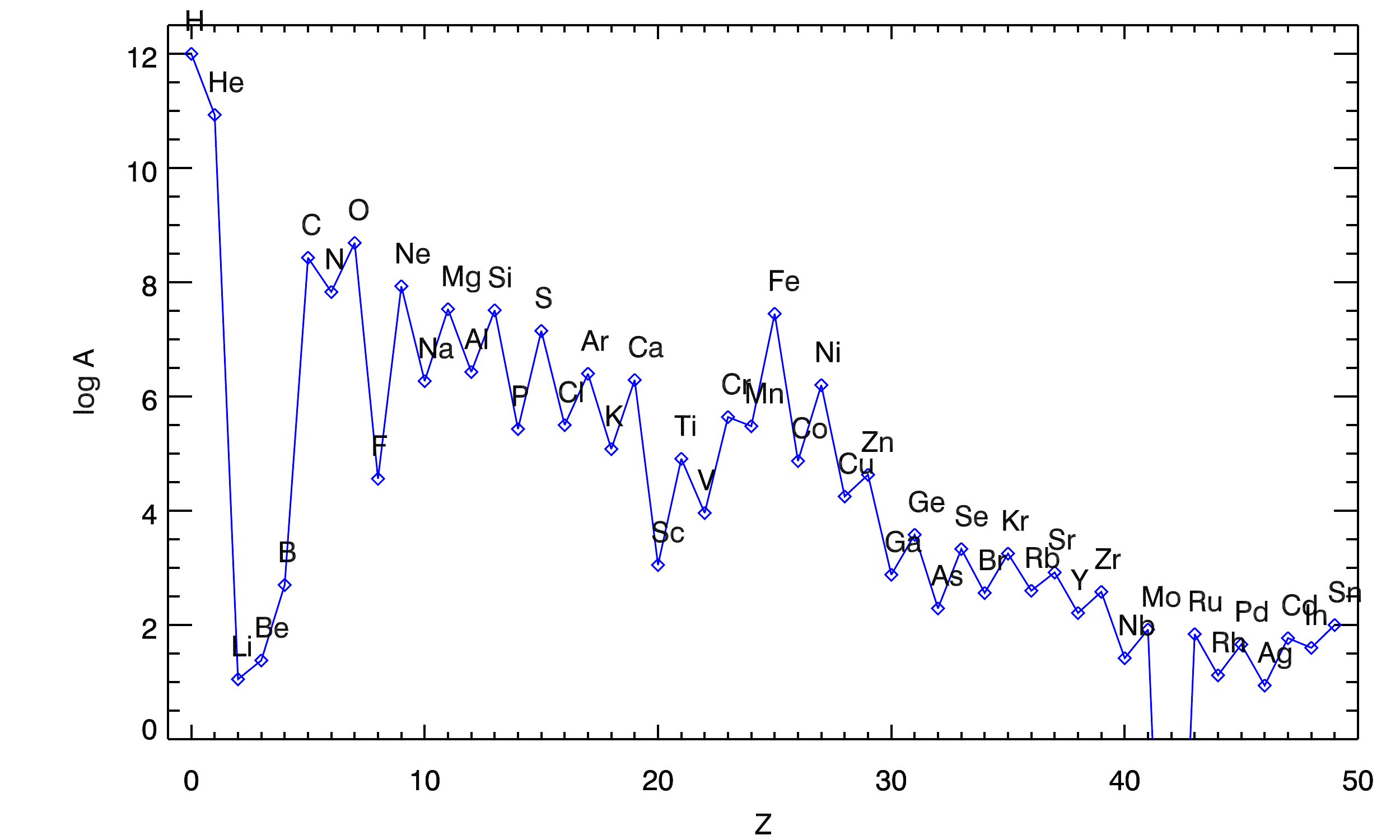}
\caption{Present-day  solar abundances, taken from AGSS09,  as a  function of
  atomic number. \label{fig:agss09}}
\end{center}
\end{figure}

Very recently, a  revision of the SSC was proposed by  \cite[][hereafter
AGSS09]{2009ARA&A..47..481A}. The new dataset (Figure~\ref{fig:agss09})
immediately became a new standard in astronomy.  But more than that, it led to
a conflict with the theory of stellar evolution thus motivating a rapid 
increase of research efforts in the field.  The predictions of standard solar
models are now in conflict with the internal structure of  the Sun, as measured
by the helioseismology. This  is known  as \textit{the solar abundance problem}
\citep{2009ApJ...705L.123S}. The problem has not been solved yet. Here we only
review the methods, recent progress in the field, and provide our opinion on the
problem.

\section{Nomenclature}
\label{sec:2}

There   are    two   commonly-used   abundance    scales:   astronomical  and
cosmo-chemical. The astronomical scale sets the 'zero' point at $\log \epsilon
(\rm H) = 12$, so then the abundance of each other element is given by:
$$
A ({\rm El}) = \log \epsilon = log (n_{\rm El}/n_{\rm H})  + 12,
$$ 
where  $n_{\rm  El}$  is   the  number  density  of  element  atomic.  The
distribution of abundances  on the  astronomical  scale, also  known as  {\it
log} scale, is  shown  in   Figure~\ref{fig:agss09}.  The cosmo-chemical  scale
normalises all abundances to the number of Si atoms, $N_{\rm  Si} = 10^6$. The
latter can be coupled to the astronomical scale through a  reference element,
usually Si because it can be easily measured in the solar spectrum and in
meteorites.

Furthermore, to compare with the models of stellar structure and evolution, it
is necessary to introduce the notations of mass fractions:
$$
X + Y + Z = 1,
$$  where $X,Y,Z$  are the  mass fractions  of H,  He, and all other heavier
elements; $Z$ being the so-called metallicity\footnote{Note, however, another
very common definition of \textit{metallicity} in stellar astrophysics, which is
the relative abundance of iron in a star relative to the Sun, $\feh = log
(n_{\rm Fe}/n_{\rm H})_{\rm star} - log (n_{\rm Fe}/n_{\rm H})_{\rm Sun}$. Both
definitions are used interchangeably, and there are transformation relations
between $Z$ and $\feh$.}.

\section{Methods}
\label{sec:3}

Different methods have been developed to determine solar abundances. They 
include empirical,  semi-empirical, and theoretical  methods.  The  former two
subclasses refer to analysis of the observed solar spectrum, from the IR and
optical (photospheric spectrum) to X-Ray  (corona), sunspots, measurements of 
the solar wind, flares, and  energetic particles. Another rich  source of 
information  is provided  by the  most primitive CI  chondritic meteorites  that
have avoided  chemical fractionation and  are  thus  believed  to  preserve  the
 solar  system pristine relative abundances of refractory metals
\citep{2009LanB...4B...44L}. Theoretical methods include inversions of
helioseismic data \citep[e.g.][]{2004ApJ...606L..85B} and nucleosynthesis models
for heavy noble gases \citep[e.g.][]{2009ARA&A..47..481A}. 
\begin{svgraybox}
The notation of photospheric abundances strictly applies only to the abundances
determined from  the spectral lines, which are  formed in the solar
'photosphere', i.e. where the dominant part of the solar radiation flux is
emitted. However,  the region is  poorly defined. Usually, lines formed at
optical depths  $-5  < \log \tau_{500}  < 0$ are tagged as photospheric, even
though the $T$ minimum occurs at  $\log \tau \sim  -3$ and above this point
chromosphere has a non-negligible influence on the line formation. The entire UV
quasi-continuum at $\lambda < 250$ nm has a chromospheric origin. But also IR
lines, as well as some strong lines in the UV and optical (e.g., Ca triplet at
$850$ nm and H$_\alpha$), may show signatures of
chromospheric emission in the cores.
\end{svgraybox}

Unfortunately, each method is prone to its own limitations and thus provides
only a  subset of data points on the element abundance diagram. So, volatile
elements: H,  He, C, N, O  and noble gases  are absent or heavily depleted in
meteorites; the solar photospheric spectrum does not contain lines of elements
with very high ionisation potential, such as He and other noble gases; some
elements have spectral lines in the wavelength regions unaccessible from the
ground, such as the B line in the far-UV. Furthermore, to convert the abundances
derived by different methods to the same scale,  a reference element is  needed.
For homogenising  meteoritic and photospheric scales,  Si is often used as the
anchor point between the two scales. Abundances from solar wind, corona, flares,
or sunspots are converted to the photospheric  scale using Ne$/$Mg or Ne$/$O 
ratios. This  involves  modelling the complex dynamical behaviour of elements
with different ionisation potentials in the outermost layers of the Sun, and the
robustness of such methods is questionable.

In short, the following methods are used for the analysis of different element
groups: 

\begin{itemize}
\item  inert He has a  very high  ionisation  potential  (24.6~eV), and  the
important  lines are  located in the far UV. Its abundance can be inferred from
the solar wind and corona, but the value is poorly-constrained and highly
variable. Theoretical models of stellar  evolution and the data from 
helioseismology  \citep{2004ApJ...606L..85B} provide a more accurate estimate,
consistent with each other to $10 \%$ (see below);
\item light elements Li, B, and  B are determined from the solar spectrum. The
important lines, are, however, model dependent: because of very low abundances
and very simple electronic configurations, the atoms give rise to one or few
spectral lines only (Li I) or they are located in a very problematic part of a
spectrum (Be II, B I). Li is depleted in the Sun by a factor of $\sim 150$
compared to meteorites, but Be and B are consistent.
\item volatile elements C, N, O can be determined from the solar coronal and
photospheric  spectrum, where they are present in the form of atomic and
molecular (e.g. CO, C$_2$, CH, OH,  NH, CN) lines. Despite the wealth of
spectral lines, the abundances of C, N, and O have always been a matter of
debate.
\item the abundances of refractory elements\footnote{We follow the definition of
refractory vs. volatile elements used in planetary sciences (not in industry),
as e.g. in \cite{2001sse..book.....T}. In this definition, a material which has
relatively high condensation temperature is refractory.} can be determined from
meteorites and from the solar photospheric spectrum. The two techniques give
generally consistent values, apart from selected elements like Co, W, Au, Pb, Pd
\citep{2009ARA&A..47..481A}. Meteoritic abundances can be directly measured in a
laboratory through chemical analysis. However, some  degree of aqueous 
alteration might  be present and samples have to be selected  carefully
\citep{2009LanB...4B...44L}. In contrast, photospheric abundances are
model-dependent, because they require a model of radiation transport in the
solar atmosphere. Moreover, there are non-negligible differences in the results
caused by different spectroscopic techniques, which are partly subjective (more
below). Redeterminations of meteoritic abundances of refractory elements have
been very robust over the years, therefore they can be taken as a reference.
\item Kr and Xe are determined from nucleosynthesis models of slow-neutron
capture process. The neutron-capture cross-section are accurately known from
experimental measurements  \citep{2009LanB...4B...44L}. 
\end{itemize}
\section{Solar abundance problem}\label{sec:4}

The solar abundance problem refers to the conflict of observed data from
helioseismology and the predictions made by stellar interior models for the Sun.
The 'problem' emerged only recently, when a new set of solar abundances computed
with the 3D and NLTE spectroscopic methods became available. When used as a
basis for calibrating the solar models, the new abundances lead to the solar
interior properties, which cannot describe the present-day Sun.

To understand the roots of this problem, we first need to delve into some
aspects of spectroscopic analysis that is the subject of the next section.

\subsection{Solar abundances and metallicity}

Traditionally, solar abundance has been derived  from the analysis of the solar
spectrum based on one-dimensional (1D) hydrostatic model atmospheres.  Moreover,
local thermodynamic equilibrium (LTE) has been generally adopted.  Both
approximations\footnote{See the lecture on 3D NLTE line formation.} were
necessary in the past because of computational limitations. An example of an SSC
based on these assumptions is that of \citep[][hereafter,
GS98]{1998sce..conf..161G}, which is in fact one of the most famous datasets in
astronomy.

Recently, stellar spectroscopy has evidenced two major developments. First it
has become possible to perform three-dimensional (3D) radiation hydrodynamics
(RHD) simulations \citep[][see for an extensive review on the
topic]{2009LRSP....6....2N}. Another major improvement is the development of
accurate non-LTE radiation transport models. The 3D RHD and NLTE models are very
successful in describing a great variety of observational data, such as
center-to-limb variation of the solar radiation field, shapes and asymmetries of
spectral lines, brightness intensity contrast, and they predict consistent
abundances for various diagnostic lines of a given chemical element. Up to date,
 the  most complete and consistent SSC set is that by
\citet{2009ARA&A..47..481A}.

\begin{figure}[b]
\begin{center}
\includegraphics[scale=.35]{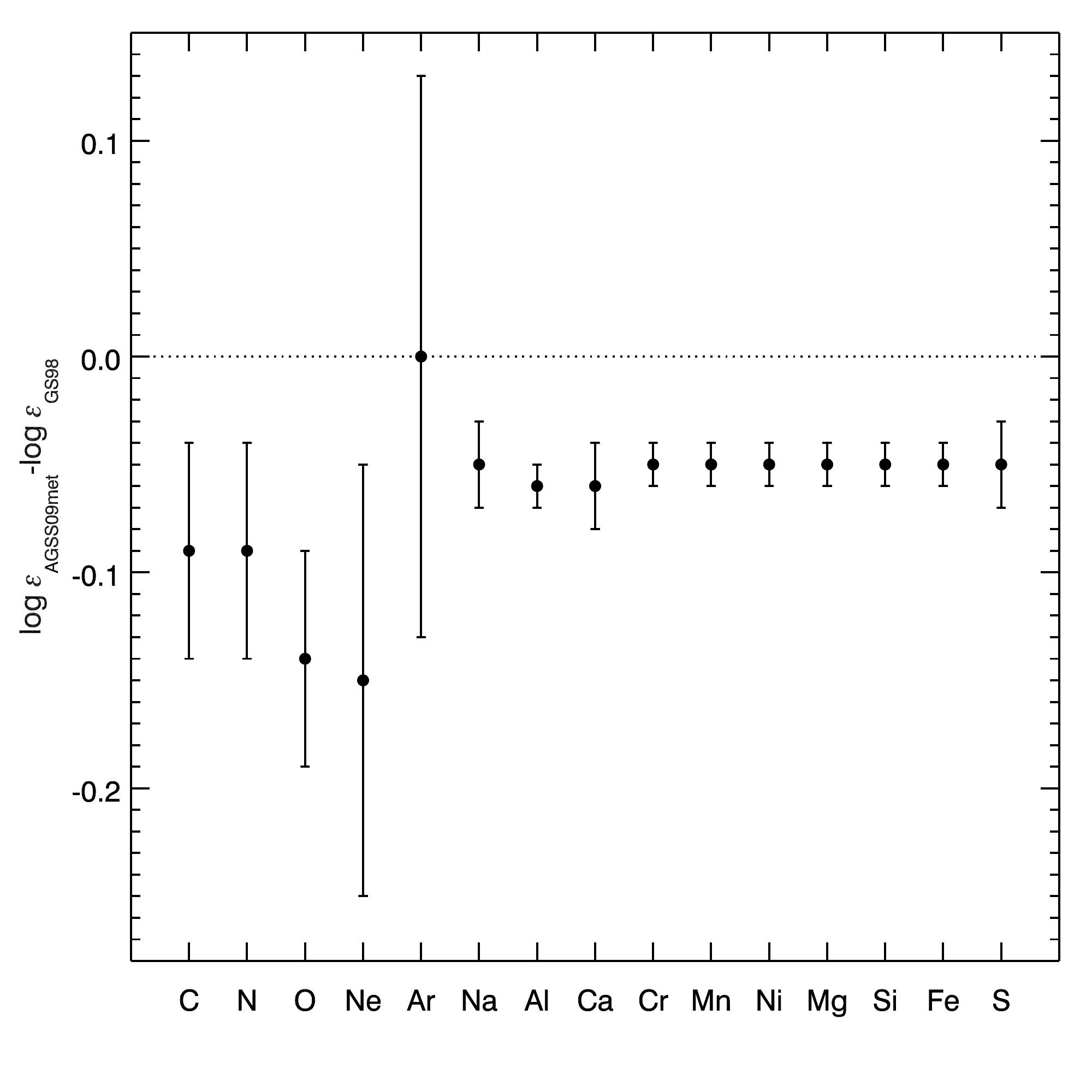}
\caption{Difference between AGSS09 and GS98 solar abundances. Elements shown
here are the most relevant for solar model calculations. We use photospheric
abundances for the volatile elements and meteoritic abundances for all other
elements in both sets of abundances. \label{fig:abddiff}}
\end{center}
\end{figure}

For the reasons given in Sec.~3, meteoritic abundances of refractory elements
and photospheric abundances for the volatile elements have been traditionally
used in solar interior models. In what follows, we adopt this combination as our
reference solar abundances, both for AGSS09 and GS98. Figure~\ref{fig:abddiff}
shows the difference between GS98 and AGSS09 results for the elements most
relevant to the solar abundance problem (C, N, O, Ne, Ar, Na, Al, Ca, Cr, Mn,
Ni, Mg, Si, Fe, S). The error bars correspond to those from  AGSS09.

AGSS09 abundances are systematically lower than the older GS98 values that can
be explained by several reasons (see a detailed discussion in AGSS09 and in
\citealt{2011CaJPh..89..327G}). First, 3D RHD models produce temperature
fluctuations in the solar atmosphere, which are absent by construction in 1D
models that results in lower molecular abundances. Secondly, it was recognised
that the important forbidden [O I] line at $630 $ nm is blended by a Ni I line,
a fact that was overlooked in previous studies and led to an overestimation of
the O abundance. Finally, the NLTE abundance obtained from the O I $777$ nm
triplet is lower compared to LTE. According to AGSS09, CNO atomic and molecular 
indicators  are now in a good agreement. Another important element, Ne, is lower
in AGSS09 because its abundance is determined from the measured Ne/O ratio in
the solar corona and the photospheric O abundance.  Finally, for the refractory
elements, the difference is caused by the lower abundance of the photospheric
determination of Si, the anchor element between the photospheric and meteoritic 
scales. The AGSS09 Si abundance is $0.05$ dex lower compared to GS98, and this
brings down all refractory abundances by the same amount.

Note that another set of CNO solar abundances has been recently provided by the
CO5BOLD collaboration \citep{2011SoPh..268..255C}. These abundances are also
based on a 3D RHD model of  the solar atmosphere, but using a different approach
abundance determinations. CO5BOLD abundances of CNO elements lie in between GS98
and  AGSS09 and the  quoted uncertainties are larger than those given in AGSS09.
The differences between AGSS09 and CO5BOLD have been attributed to the
differences in the spectrum normalisation and the choice of diagnostics lines
\citep{2011CaJPh..89..327G, 2012ASPC..462...41G}, which are both partly
subjective aspects of a spectroscopic analysis.

In summary, the differences between GS98 and  AGSS09 abundances amount  to  20\%
to  40\%  for  CNO, Ne and 12\% for refractories. Since spectroscopy provides
only the relative abundances of metals to hydrogen, it is very convenient to
combine all these numbers into the metal-to-hydrogen mass ratio ${\rm 
\left(Z/X\right)_\odot}$.  For  the three sets of SSC discussed above, the
values\footnote{Note that the given $\left(Z/X \right)$ were computed using
photospheric abundances for the volatile elements and meteoritic abundances for
all other elements. Therefore, it is slightly different from the present-day
photospheric value, as e.g. given by \citep[][Table 4]{2009ARA&A..47..481A},
$\left(Z/X \right) = 0.0181$.} are:
$$
{\rm \left(Z/X \right)_{\rm GS98}= 0.0229; \, \, \, \left(Z/X \right)_{\rm
CO5BOLD}=
0.0209; \, \, \, \left(Z/X \right)_{\rm AGSS09}= 0.0178.}
$$

${\rm Z/X}$ is one of the three fundamental constraints that have to be
satisfied by SSMs. 
The value of ${\rm Z/X}$ provided by CO5BOLD is obtained by complementing their
photospheric measurements with abundances from Lodders et al. (2009) for
refractories and noble gases.

\begin{svgraybox}
SSMs  are one-dimensional evolutionary  models  of  a 1~${\rm  M_\odot}$  star,
starting from a homogeneous model in the pre-main sequence up to the present-day
age of the solar system  $\tau_\odot=4.57$~Gyr.  At  this age, the  model has 
to satisfy three observational constraints: the  present-day luminosity and
radius (${\rm L_\odot=  3.8418\times  10^{33}\,\mbox{ers  s$^{-1}$}}$ and  ${\rm
 R_\odot= 6.9598\times10^{10}\,\mbox{cm}}$)  and  the $\zxo$. Three free
parameters are calibrated to fulfil these conditions: the mixing length
parameter $\alpha_{\rm MLT}$ that controls the efficiency of convection in the
Mixing Length Theory, ${\rm Y_{ini}}$  and  ${\rm  Z_{ini}}$. The relative
abundances of individual metals are assumed to be the same for a given $\zxo$.
The initial hydrogen abundance  is determined from the normalisation ${\rm
X+Y+Z=1}$.
\end{svgraybox}

\subsection{Helioseismology and the Standard Solar Models}

From the discussion above, it is clear that solar abundances are critical for
the calibration of SSM. Here, we focus our discussion on the GS98 and AGSS09 
abundances, and label the solar models accordingly, i.e. GS98 SSM and AGSS09
SSM.

The internal structure of the Sun can be accurately determined by
helioseismology. The observed  oscillation spectrum can be derived from the
measured light curves or from the Doppler shifts of photospheric absorption
lines. The quantities we are interested in are  the trapped 'eigenmodes', i.e.
standing waves. The resonant cavity of these modes has its outer boundary in the
solar atmosphere but the location of the inner boundary depends on the
characteristics of  each individual mode: its frequency and angular degree. As a
result, different modes map the Sun to different depths and that allows to
determine the physical properties of the interior structure as a function of
depth, all the way down to the core (see \citealt{2002RvMP...74.1073C} for a 
comprehensive  review  on helioseismology). Several very important physical
characteristics of the Sun can be derived: the radial profile of the sound speed
and the density in the interior, the location of the base of the convective
envelope $\rcz$, and the helium abundance of the envelope $\ys$. 

\begin{figure}[b]
\begin{center}
\includegraphics[scale=.32]{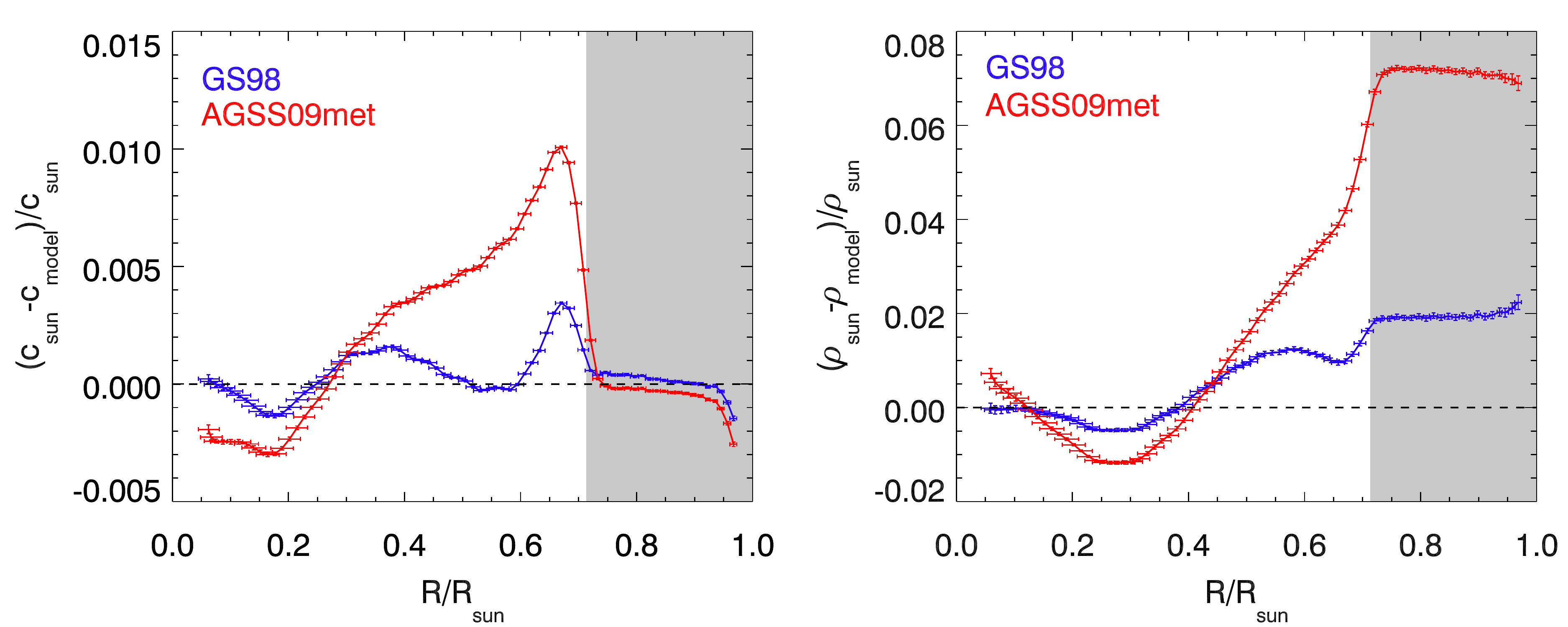}
\caption{Profiles of the relative difference in sound speed (left panel) and
density (right panel)  between the Sun and two SSMs. Models  are labeled
according to  the SSC  that was  adopted in the calibration of the SSM. The 
grey area denotes  the solar  convective  envelope. The error  bars reflect the
uncertainties from the helioseismic data. \label{fig:ssmhelio}} 
\end{center}
\end{figure}

The main results for the SSMs and the results from helioseismology are presented
in Table~\ref{tab:ssm}. SSMs results are taken from \cite{2011ApJ...743...24S} 
but, with small differences, they are common to SSM calculations from other
authors \citep{2004ESASP.559..574M, 2006ApJ...649..529D, 2010ApJ...713.1108G}.
$\dc$  and $\drho$  are the average root-mean-square deviations of the relative
difference between the model (SSM) and the solar (helioseismic) quantities. Note
that the errors for the helioseismic $\rcz$  and $\ys$ are extremely small.
Figure~\ref{fig:ssmhelio} also shows the relative differences of the sound speed
profile and of the mean density profile for both models.
\begin{table}
\begin{tabular}{lccccccccc}
\hline
& $\zxo$ & $Z_{\odot}$ & $Y_{\rm S}$ & $R_{\rm CZ}/{\rm R_\odot}$ & $\dc$ &
$\drho$ & $Z_{\rm ini}$ & $Y_{\rm ini}$ \\
\hline 
GS98 & 0.0229 & 0.0170 & 0.243 & 0.712 & 0.0009 & 0.011 & 0.0187 & 0.272 \\
AGSS09 & 0.0178 & 0.0134 & 0.232 & 0.723 & 0.0037 & 0.040 & 0.0149 & 0.262 \\
\hline
Solar   &$0.0229/0.0178^{\rm   a}\  \   $&   $0.0168/0.0131^{\rm   a}\  \   $&
$\ 0.2485^{\rm b}$ & $\ 0.713^{\rm b}$ & 
0(def) & 0(def) & --- & --- \\
& & & $\pm 0.0035$ & $\pm0.001$ \\
\hline
\end{tabular}
\caption{Main  characteristics of the standard  solar  models  and comparison 
to the solar  (helioseismic) results.  $^{\rm a}$Refers to  GS98 and  AGSS09
SSCs, respectively; $^{\rm b}$Basu \& Antia (2004). \label{tab:ssm} }
\end{table}

First of all, the GS98 SSM shows a very good agreement with the helioseismic
inferences for $\ys$  and $\rcz$ (Table~\ref{tab:ssm}). The results obtained
with the AGSS09 composition are in stark contrast with the latter. Furthermore,
the choice of $\zxo$ has a direct impact on the calibration of solar models, as
seen from the differences in the initial mass fractions of helium and of metals,
${\rm Y_{ini}}$ and  ${\rm Z_{ini}}$. The  changes in ${\rm Z_{ini}}$ almost
directly reflect the differences in $\zxo$. Metals are the dominant contributors
to the radiative opacity $\kappa$ in the solar interior, which in turn
determines the temperature gradient in the radiative region  (white area,
Figure~\ref{fig:ssmhelio}). A lower metallicity leads to a  smaller temperature
gradient in this region and, by virtue of the Schwarzschild  convection 
criterion, a  shallower depth of the convective envelope $\rcz$
(Table~\ref{tab:ssm}). In the convective envelope (shaded area,
Figure~\ref{fig:ssmhelio}), where the temperature gradient does not depend on
$\kappa$, but only on the equation of state, both the GS98 and AGSS09 models
agree well with the seismic data, $\delta c /c \sim 0$. However, in the
radiative zone, where $\kappa$  affects the solar structure, differences in the
sound speed profiles show up.

The density values at different solar radii are strongly correlated because the
density profile is constrained by the total mass of the Sun. The large
differences seen in the convective envelope are anticorrelated with changes in
the deeper interior where density is larger (region between 0.15 and 0.4~${\rm
R_\odot}$). A smaller difference between the Sun and the model in a deep region
leads to a large, compensating, difference in outer layers. Finally, the surface
He abundance $\ys$ is also affected by the metallicity of the models. The reason
is that SSMs are constrained by ${\rm L_\odot}$. The AGSS09 SSM, with its
shallower temperature gradient, has a lower core temperature, leading to a
slower rate of nuclear energy generation.  But, because nuclear reactions are
the only relevant  energy  source  in the Sun, the decrease in temperature has
to be compensated by another means so that the fusion of hydrogen still produces
energy at the same rate. This is achieved  in the AGSS09 SSM by the increase of
hydrogen abundance or, equivalently, by the decrease of helium abundance. The
lower  $\ys$ in  the AGSS09 SSM also conflicts with helioseismology.

In summary, all manifestations of the solar abundance problem can be traced back
to the  lower radiative opacity in the interior that is a consequence of lower
metal  abundances in the AGSS09 dataset. Oxygen, neon, and iron are very
important in this respect, as they contribute to  $\kappa$ in the region around
$\rcz$, where their fractional contribution to $\kappa$ is about $25\%$,
$15~\%$, and $10\%$ respectively (\citealt[][Fig. 12]{2008PhR...457..217B},
\citealt[][Fig. 10]{2013arXiv1312.3885V} Other abundant refractories like
magnesium and silicon are less relevant (opacity contribution of $<4\%$).

\section{Possible solutions}
\label{sec:5}

The implications of the solar abundance problem in the astrophysical context
are, in fact, very large. The problem is that stellar evolution models, when
applied to the Sun, produce results which are incompatible with observations
(helioseismology), if these models adopt the solar chemical composition obtained
by the state-of-the-art spectroscopic models. However, both stellar evolution
and stellar atmosphere theory are general. All models based on these theories
are calibrated on the Sun and they are routinely used to interpret any other
star or stellar population. Therefore, our understanding of stars and galaxies
in general is nowhere better than our current knowledge of the Sun.

What are the possible ways to reconcile the spectroscopic measurements with the
solar interior models? Here we review the key possibilities that can be or have
already been considered.

\begin{itemize}

\item Model  atmospheres. It  is difficult to  assess and  quantify
uncertainties in  stellar atmosphere models. Possible sources of  uncertainties
are: numerics (e.g., numerical methods, resolution of simulations),  accuracy of
the input physics  (e.g., the equation  of  state), approximate treatment of
physical processes (e.g. simplified radiative transfer). The models can be
tested by comparison with observations and by comparing models from different
groups. For example, \citet{2012A&A...539A.121B} showed that the average
stratifications of different 3D hydrodynamical model atmospheres agree well with
each other. The 3D hydro models are also much more successful than classical 1D
static models in reproducing a wealth of observational information, including
the line shapes, center-to-limb variation, brightness contrast
\citep{2009ARA&A..47..481A}. However, there are implicit assumptions, which
remain to be tested.
\item The spectroscopic analysis. This is also a highly non-trivial problem: 
the uncertainties in the atomic data, line broadening, the continuum placement,
selection of lines, directly impact the calculated abundances.
\end{itemize}

NLTE radiative transfer and 3D hydrodynamical model atmospheres are clearly
setting a new basis for spectroscopy, however, it is still very difficult to
combine them in one framework. 3D NLTE calculations can be performed on
realistic timescales only for the simplest atoms, such as Li and O. More complex
atoms can be only treated in a very approximated form (see the lecture on 3D
NLTE modelling). Future developments related to 3D and NLTE may eventually lead
to a revisions of abundances. However, it is unclear whether the abundances will
increase back to the level needed for the solar abundance problem to disappear. 

From the perspective of solar models, a number of possibilities to solve the
solar abundance crisis have also been considered.

\begin{itemize}

\item The accuracy of radiative opacities for stellar interior models could be
questioned. Recently, Villante et al. (2013) have shown that current
helioseismic and solar neutrino data constrain well the opacity profile of the
Sun, independently of the reference solar models and abundances. But in models,
the opacity profile results from a combination of atomic opacity calculations
and a given solar abundance. The effects of a decreasing metallicity can be
mimicked by an increase in opacity. In fact, it has been shown that an increase
in the radiative opacities in the range of 15  to 20\% at  the base of the
convective zone,  smoothly decreasing to 3 to 4\% in the solar core, would
suffice to reconcile AGSS09 composition with the helioseismic  results
\citep{2004ApJ...614..464B, 2009A&A...494..205C, 2010ApJ...724...98V}. This 
solution is very attractive, because radiative  opacities are the result of very
sophisticated  (and, unfortunately, incomplete) theoretical calculations of
interaction of atoms and radiation in extremely dense physical environments.
There is basically no experimental data to support these calculations. In
contrast to the atmosphere models, presently the best that can be done to gain
confidence in such calculations is to compare the results from  different 
groups. Three opacity sets have been compared by \cite{2012ApJ...745...10B}, who
found much more modest differences amounting to just 3\% at the base of the
convective zone, much smaller than needed. A possible way out of problem imposed
by the degeneracy between opacity and composition might be offered by solar
neutrino measurements. In particular, the neutrino fluxes originating in the
CN-cycle, depend linearly with the C and N abundance in the solar core. But C
and N do not affect the solar opacity, so if the CN neutrino fluxes are measured
(e.g. by the Borexino Experiment or SNO+) the solar core C and N abundance can
be determined independently of the solar opacity \citep{2013PhRvD..87d3001S}.

\item Enhanced gravitational settling in the Sun. In the conditions of the solar
interior, chemical elements suffer a slow segregation process due to the
combined effect of the  gravitational and electric field. This  process, 
generally known as gravitational settling affects differently elements
(actually, isotopes) based on their nuclear charge to  mass ratio ${\rm 
Z_{nuc}/A_{nuc}}$. However, in the Sun, settling rates are quite similar for all
metals and helium \citep{1998ApJ...504..539T}. SSM calculations show that the
gravitational settling  has  led to the $\sim 10-12 \%$ decrease of the solar
surface metallicity and helium abundance  with  respect  to  the  primordial 
values  (compare the initial and surface values in Table~\ref{tab:ssm}).
Increasing the efficiency of settling, one could construct a solar model with
the initial composition comparable to GS98 and a low metallicity in the
convective envelope, compatible with AGSS09. However, such a model would predict
a too low $\ys$, worsening the agreement with helioseismology, as discussed, for
example, in \cite{2010ApJ...713.1108G}. They do not  offer a satisfactory
solution. In order to improve the agreement with helioseismology, an ad-hoc
modification of settling rates should be applied such that metals  sink faster,
but helium slower. Such an ad-hoc solution is not sufficiently justified and
should be avoided.

\item  Solar models  with the accretion of  metal-poor  material. Young stars 
interact and accrete material from their proto-planetary disk (see Williams \&
Cieza 2011 for a comprehensive review). In the inner solar system, planets are
clearly metal-rich compared to the Sun. This is true for Jupiter as well. The
process of  planet  formation is  likely  to  alter  the  average  composition
of  the proto-planetary disk. If a part of the disk, partially depleted in
metals, is accreted onto the young Sun, the solar interior will have a higher
metal content  than the envelope. Such models have been considered by
\citet{2007A&A...463..755C}, \citet{2010ApJ...713.1108G} and in more
detail by \citet{2011ApJ...743...24S}. Unfortunately, only partial solutions to
the problem have been found. One can fine-tune the mass and chemical composition
of the accreted material so that $\rcz$ is close to the seismic value but, at
the expense of $\ys$. Under some conditions, $\ys$ agreement can be improved,
but at the expense of degrading $\rcz$.

\item Enhanced solar neon abundance. As discussed before, the determination of
the solar neon abundance is indirect. The coronal Ne/O ratio can be measured
and,  by assuming the same ratio is present in  the solar photosphere, the
photospheric neon abundance can be determined. Keep in mind this  is a strong
assumption.  How different elements  are transported  from stellar photospheres 
to corona is far from being well understood, and this also depends strongly on
other issues such as the stellar  activity \citep{2008A&A...486..995R}.  As a
consequence,  the Ne/O  coronal  ratio  is not  constant in time (neither for
the  Sun nor for other stars). 

Ne is an interesting element for SSMs because it contributes to the radiative
opacity at the  base of  the convective envelope. \citet{2005ApJ...631.1281B}
and  \citet{2005ApJ...620L.129A} constructed solar models with arbitrarily
enhanced neon abundance and found that an increase of a factor of about 2 would
be necessary to solve the  solar abundance problem. \citet{2005Natur.436..525D}
found, based on X-ray spectroscopy of nearby active stars, that neon abundances
were a factor of $2.5$ larger with respect to oxygen in those stars  with 
respect to  the measured  solar value and  concluded that  the  solar value was
underestimated. However, it is now thought that the large Ne/O values observed
in the coronae of very active stars are linked to the high activity levels and
do not reflect the photospheric Ne$/$O ratio \citep{2008A&A...486..995R}. Thus
it does not seem likely that the  solar Ne/O could be large enough to solve the
solar abundance problem.

\item Non standard solar models (non-SSM). Undoubtedly the  SSM, with all its  
intricacies, is a simplified   picture  of  the  actual  Sun and  its evolution.
Physical  processes such as rotation or  internal magnetic fields are not
accounted for in the  SSM, and  effects such  as the  transport of angular
momentum in the solar interior might have measurable consequences in the solar
structure. Modelling these processes is an inherently multi-dimensional  problem
and such models (to study not just the present-day structure, but the evolution
of the course of $4.57~$ Gyr) have  not yet been  constructed. The models are
also not  feasible with present-day computational   capabilities.   Simplified
prescriptions  have  been implemented into 1D solar models. These  parametrized
models present a number of problems:  first they contain new free parameters 
that have to be tuned to reproduce various observables (e.g. the  internal
rotation profile of the  Sun), losing their predictive power. Second, such
models based on the GS98 composition give an overall description of the solar
structure that is worse than SSM results using  AGSS09 \citep[see
e.g.][]{2010ApJ...715.1539T}. So, for non-SSM, the problem lies at a much deeper
level:  it is not yet possible to include realistic description of dynamical
processes in the solar interior.

\end{itemize}

\section{Epilogue}

The solar abundance problem is that stellar evolution models, when
applied to the Sun, produce results which are incompatible with observations
(helioseismology), if these models adopt the solar chemical composition
obtained by the state-of- the-art spectroscopic models. This should be taken as
a manifestation of our incomplete understanding of the theories of stellar
atmospheres and/or  of stellar structure and evolution.

As discussed in these notes, at present there is no clear solution to the
problem. A plausible solution is that radiative opacity calculations for stellar
interiors are off by 3 to 20\% or so. While only 3\% is the maximum discrepancy
that has been found among different opacity calculations, changes of up to 
20\% might not be unreasonable. For example, the OPAL opacities
\citep{1992ApJ...401..361R}, that first saw the light back in 1992, implied
changes of up to a factor of $3$ with respect to previously available
calculations.

Continuous progress in different subfields might lead to a revision of
the solar abundances and to changes in the solar interior models. Various
options remain to be studied in greater detail and with more powerful computers,
including radiative transfer models in the solar atmosphere, opacity
calculations and laboratory experiments, inclusion of more realistic physics in
the stellar interior modelling. It is up to you, new generation of scientists,
to find the ultimate solution to the problem.

\begin{acknowledgement}
Figure 3 reproduced by permission of the AAS: Serenelli et al. 2011,
ApJ, 743:24. AMS is supported by the MICINN grant AYA2011-24704. This work was
partly supported by the European Union FP7 programme through ERC grant number
320360.
\end{acknowledgement}

\bibliographystyle{spbasic}
\bibliography{references}

\end{document}